\documentclass[%
reprint,
superscriptaddress,
 amsmath,amssymb,
 aps,
prc,
floatfix,
nofootinbib 
]{revtex4-2}

\usepackage{multirow,booktabs,siunitx}

\usepackage{graphicx}
\usepackage{dcolumn}
\usepackage{bm}
\usepackage{soul}

\usepackage{natbib} 
\DeclareMathOperator{\sgn}{sgn}

\usepackage[dvipsnames]{xcolor}
\usepackage{orcidlink}

\makeatletter
\let\@fnsymbol\@alph 
\makeatother

\begin{document}



\author{C.~Abel}
 \affiliation{Department of Physics and Astronomy, University of Sussex, Falmer, Brighton BN1 9QH, United Kingdom}

\author{K.~Bodek}
  \affiliation{Marian Smoluchowski Institute of Physics, Jagiellonian University, 30-348 Cracow, Poland}

\author{E.~Chanel}
 \thanks{Corresponding author: chanele@ill.fr}
 \altaffiliation[Present address: ]{Institut Laue-Langevin, CS 20156 F-38042 Grenoble Cedex 9, France}
 \affiliation{Laboratory for High Energy Physics and Albert Einstein Center for Fundamental Physics, University of Bern,
CH-3012 Bern, Switzerland}

\author{P.-J.~Chiu~\orcidlink{0000-0002-3772-0090}}
 \altaffiliation[Present address: ]{Department of Physics, National Taiwan University, 10617 Taipei, Taiwan}
 \affiliation{ETH Zürich, Institute for Particle Physics and Astrophysics, CH-8093 Zürich, Switzerland}
  \affiliation{Paul Scherrer Institut, CH-5232 Villigen PSI, Switzerland}

\author{C.B.~Crawford}
  \affiliation{Department of Physics and Astronomy, University of Kentucky, Lexington, KY 40504, USA}


\author{M.~Daum}
 \affiliation{Paul Scherrer Institut, CH-5232 Villigen PSI, Switzerland}

 \author{C.B.~Doorenbos}
  \affiliation{ETH Zürich, Institute for Particle Physics and Astrophysics, CH-8093 Zürich, Switzerland}
  \affiliation{Paul Scherrer Institut, CH-5232 Villigen PSI, Switzerland}

\author{S.~Emmenegger}
 \affiliation{ETH Zürich, Institute for Particle Physics and Astrophysics, CH-8093 Zürich, Switzerland}

\author{M.~Fertl~\orcidlink{0000-0002-1925-2553}}
  \affiliation{Institute of Physics, Johannes Gutenberg University Mainz, 55128 Mainz, Germany}
  
\author{P.~Flaux}
  \affiliation{Université de Caen Normandie, ENSICAEN, CNRS/IN2P3, LPC Caen, UMR6534, 14000 Caen, France}
  
\author{A.~Fratangelo~\orcidlink{0000-0001-9964-601X}}
  \altaffiliation[Present address: ]{Los Alamos National Laboratory, Los Alamos, NM 87545, USA}
  \affiliation{Laboratory for High Energy Physics and Albert Einstein Center for Fundamental Physics, University of Bern,
CH-3012 Bern, Switzerland}

\author{W.C.~Griffith~\orcidlink{0000-0002-0260-1956}}
 \affiliation{Department of Physics and Astronomy, University of Sussex, Falmer, Brighton BN1 9QH, United Kingdom}
 
\author{P. Harris}
 \affiliation{Department of Physics and Astronomy, University of Sussex, Falmer, Brighton BN1 9QH, United Kingdom}

 \author{K.~Kirch~\orcidlink{0000-0002-1720-7636}}
 \affiliation{ETH Zürich, Institute for Particle Physics and Astrophysics, CH-8093 Zürich, Switzerland}
  \affiliation{Paul Scherrer Institut, CH-5232 Villigen PSI, Switzerland}

\author{V.~Kletzl}
 \affiliation{ETH Zürich, Institute for Particle Physics and Astrophysics, CH-8093 Zürich, Switzerland}
 \affiliation{Paul Scherrer Institut, CH-5232 Villigen PSI, Switzerland}

\author{P.A.~Koss}
  \altaffiliation[Present address: ]{Fraunhofer Institute for Physical Measurement Techniques, 79110 Freiburg, Germany}
  \affiliation{Instituut voor Kern- en Stralingsfysica, University of Leuven, B-3001 Leuven, Belgium}

\author{J.~Krempel}
 \affiliation{ETH Zürich, Institute for Particle Physics and Astrophysics, CH-8093 Zürich, Switzerland}

\author{B.~Lauss~\orcidlink{0000-0002-1986-391X}}
 \affiliation{Paul Scherrer Institut, CH-5232 Villigen PSI, Switzerland}

\author{T.~Lefort~\orcidlink{0000-0003-2198-2093}}
\affiliation{Université de Caen Normandie, ENSICAEN, CNRS/IN2P3, LPC Caen, UMR6534, 14000 Caen, France}

\author{P.~Mohanmurthy~\orcidlink{0000-0002-7573-7010}}
 \altaffiliation[Present address: ]{Laboratory for Nuclear Science, MIT, 77 Mass. Ave., Cambridge, MA 02139, USA}
 \affiliation{ETH Zürich, Institute for Particle Physics and Astrophysics, CH-8093 Zürich, Switzerland}
 \affiliation{Paul Scherrer Institut, CH-5232 Villigen PSI, Switzerland}

\author{O.~Naviliat-Cuncic}
  \affiliation{Université de Caen Normandie, ENSICAEN, CNRS/IN2P3, LPC Caen, UMR6534, 14000 Caen, France}

\author{D.~Pais}
 \affiliation{ETH Zürich, Institute for Particle Physics and Astrophysics, CH-8093 Zürich, Switzerland}
 \affiliation{Paul Scherrer Institut, CH-5232 Villigen PSI, Switzerland}
  
\author{F.M.~Piegsa~\orcidlink{0000-0002-4393-1054}}
 \thanks{Corresponding author: florian.piegsa@unibe.ch}
  \affiliation{Laboratory for High Energy Physics and Albert Einstein Center for Fundamental Physics, University of Bern,
CH-3012 Bern, Switzerland}

\author{G.~Pignol~\orcidlink{0000-0001-7086-0100}}
  \affiliation{Université Grenoble Alpes, CNRS, Grenoble INP, LPSC-IN2P3, 38026 Grenoble, France}

\author{C.~Pistillo~\orcidlink{0000-0001-8131-9440}}
  \affiliation{Laboratory for High Energy Physics and Albert Einstein Center for Fundamental Physics, University of Bern,
CH-3012 Bern, Switzerland}

\author{D.~Rebreyend}
  \affiliation{Université Grenoble Alpes, CNRS, Grenoble INP, LPSC-IN2P3, 38026 Grenoble, France} 
  
\author{I.~Rienaecker}
  \affiliation{ETH Zürich, Institute for Particle Physics and Astrophysics, CH-8093 Zürich, Switzerland}
 \affiliation{Paul Scherrer Institut, CH-5232 Villigen PSI, Switzerland}

\author{D.~Ries}
\affiliation{Paul Scherrer Institut, CH-5232 Villigen PSI, Switzerland}

\author{S.~Roccia~\orcidlink{0009-0004-4752-5442}}
  \affiliation{Université Grenoble Alpes, CNRS, Grenoble INP, LPSC-IN2P3, 38026 Grenoble, France} 

\author{D.~Rozpedzik}
  \affiliation{Marian Smoluchowski Institute of Physics, Jagiellonian University, 30-348 Cracow, Poland}
  
\author{P.~Schmidt-Wellenburg \orcidlink{0000-0001-5474-672X}}
 \affiliation{Paul Scherrer Institut, CH-5232 Villigen PSI, Switzerland}

\author{N.~Severijns}
  \affiliation{Instituut voor Kern- en Stralingsfysica, University of Leuven, B-3001 Leuven, Belgium}

  \author{K.~Svirina}
 \altaffiliation[Present address: ]{Institut Laue-Langevin, CS 20156 F-38042 Grenoble Cedex 9, France}
  \affiliation{Université Grenoble Alpes, CNRS, Grenoble INP, LPSC-IN2P3, 38026 Grenoble, France} 
  
\author{J.~Thorne~\orcidlink{0000-0002-3905-5549}}
  \affiliation{Laboratory for High Energy Physics and Albert Einstein Center for Fundamental Physics, University of Bern,
CH-3012 Bern, Switzerland}

\author{S.~Touati}
  \affiliation{Université Grenoble Alpes, CNRS, Grenoble INP, LPSC-IN2P3, 38026 Grenoble, France}

\author{E.~Wursten}
  \altaffiliation[Present address: ]{Centre for Cold Matter, Imperial College London, London, SW7 2AZ, United Kingdom}
  \affiliation{Instituut voor Kern- en Stralingsfysica, University of Leuven, B-3001 Leuven, Belgium}

\author{N.~Yazdandoost~\orcidlink{0000-0001-6768-795X}}
 \altaffiliation[Present address: ]{TRIUMF, Vancouver, British Columbia V6T 2A3, Canada}
  \affiliation{Paul Scherrer Institut, CH-5232 Villigen PSI, Switzerland}  \affiliation{Department of Chemistry - TRIGA site, Johannes Gutenberg University Mainz, 55128 Mainz, Germany}

\author{J.~Zejma}
  \affiliation{Marian Smoluchowski Institute of Physics, Jagiellonian University, 30-348 Cracow, Poland}

\collaboration{nEDM Collaboration}\noaffiliation

\date{\today}

\title{Measurement of the neutron incoherent scattering length of \texorpdfstring{${}^{199}$Hg}{199Hg} using stored ultracold neutrons}

\begin{abstract}


We present the first direct measurement of the neutron incoherent scattering length of ${}^{199}$Hg. The measurement was performed with the Ramsey apparatus of the neutron electric dipole moment experiment located at the Paul Scherrer Institute. The incoherent scattering length $b_\textrm{i}$ was determined by investigating the pseudo-magnetic effect due to the strong interaction between the neutron spins and the nuclear spins of mercury atoms. The resulting frequency shift of the neutron Larmor precession frequency was determined for various ${}^{199}$Hg density and polarization values. 
The obtained value of $b_\textrm{i}= (-16.2 \pm 2.0)\,$fm agrees in magnitude with previous determinations and provides the so-far unknown sign of the quantity.


\end{abstract}



\maketitle


\section{Introduction}

Neutrons interact with an atomic nucleus via the strong force.
The range of the neutron-nucleus interaction is much smaller than the de Broglie wavelength of ultracold, cold, or thermal neutrons. Hence, at low energies, the scattering is isotropic and can be described with a single parameter, namely, the bound scattering length $b$.
The scattering length varies by isotope and by element in a way that appears random and the value may either be positive or negative.
 For historical reasons, its spin-independent and spin-dependent components are called coherent and incoherent scattering lengths, respectively.
 In 1947, Fermi and Marshall published the first set of neutron coherent scattering lengths measurements with nuclei \cite{PhysRev.71.666}.  
Since then, further measurements have been performed and extensive tables of scattering lengths summarize the results \cite{Sears1992, KOESTER}.
In 1965, Baryshevskii and Podgortskii expressed the idea of a pseudo-magnetic field experienced by the neutrons in a target of polarized nuclear spins in order to describe the spin-dependent interaction \cite{PseudoMag}. 
A few years later, Abragam et al.\ started a series of incoherent scattering length measurements using a neutron Ramsey apparatus sensitive to such pseudo-magnetic fields \cite{IncMeasRams}.
Later, similar incoherent scattering lengths measurements have been carried out for various nuclides by different research groups \cite{Roubeau1974,Abragam1975,Glattli1979,MALINOVSKI1981103,zimmer2002,Piegsa2008b,HUBER2009235,lu2023measurement,VANDENBRANDT2009231,abragam1982nuclear}.\\
In the realm of fundamental neutron physics and more specifically neutron electric dipole moment (EDM) searches, spin polarized gases, for instance of mercury and xenon, often serve as sensitive atomic magnetometer systems.
In these experiments, the low-density gases co-inhabit the same volume as storable ultracold neutrons (UCN) sampling simultaneously the same magnetic field environment \cite{BAKER2014184,TriumfXecoMag,Ramsey1984}.
Here, we focus on the co-magnetometer isotope of $^{199}$Hg, which is of particular interest for the nEDM collaboration with the experiment located at the Paul Scherrer Institute \cite{Ayres2021}.
The reason being that a potential correlation between a pseudo-magnetic field resulting from its incoherent scattering length $b_\textrm{i}$ and a reversal of the electric field could mimic a non-zero electric dipole moment. 
This was considered during the analysis of the current best limit, but due to the unknown sign of $b_\textrm{i}$, the result could not be corrected for this systematic effect  \cite{EDM2020}. 
Nevertheless, the knowledge of its absolute value provided in the literature of $|b_{\textrm{i},lit}| = (15.5 \pm 0.8)\,\textrm{fm}$ allowed an estimation of the magnitude of this effect \cite{Sears1992}. It led to a significant contribution to the systematic uncertainty of $7\times 10^{-28}\,e\,$cm, which represents more than one-third of the total systematic uncertainty budget of $2\times 10^{-27}\,e\,$cm of the experiment. 
In this article, we present the first direct measurement of the neutron incoherent scattering length of ${}^{199}$Hg using UCN and the neutron Ramsey apparatus of the nEDM experiment performed in 2017 \citep{apparatusPSI,ChanelThesis,PhysRevD.92.092003}. 


\section{Incoherent scattering length}

The strong interaction between a neutron and a bound nucleus is described by the bound scattering length
\begin{equation}
\begin{aligned}
b=b_{\textrm{c}} + \frac{2 b_{\textrm{i}}}{\sqrt{I(I+1)}}\vec{s}\cdot\vec{I}.
\end{aligned}
\label{BoundScatt}
\end{equation}
It is composed of the coherent spin-independent scattering length $b_{\textrm{c}}$, and a spin-dependent contribution. The latter is given by the scalar product of the neutron spin $\vec{s}$ with the spin of the nucleus $\vec{I}$, and the incoherent scattering length $b_{\textrm{i}}$ \cite{Sears1992}.
The interaction of neutron spins with a homogeneous sample of polarized nuclei can be described with a spin-dependent Fermi potential which is proportional to $b_{\textrm{i}}$ of the polarized nuclear species
\begin{equation}
\begin{aligned}
V_{\textrm{F},\textrm{i}}=  \frac{4 \pi \hbar^2}{m_{\textrm{n}}} \, b_{\textrm{i}} \,\rho \sqrt{\frac{I}{I+1}} \vec{s}\cdot\vec{P},
\end{aligned}
\label{eq.VFS}
\end{equation}
where $\hbar$ is the reduced Planck's constant, $\rho$ the nuclear number density, $m_{\textrm{n}}$ the mass of the neutron, and $\vec{P}=\langle \vec{I} \rangle / I$ the polarization vector averaged over all nuclear spins. 
In turn, such a potential can  be interpreted as a pseudo-magnetic field 
\begin{equation}
\begin{aligned}
\vec{B}^{*}= - \frac{4 \pi \hbar}{m_{\textrm{n}}  \gamma_{\textrm{n}} } \, b_{\textrm{i}}  \,\rho \sqrt{\frac{I}{I+1}} \vec{P},
\end{aligned}
\label{eq.pseudoB}
\end{equation}
 using the formula for a magnetic potential $V=-\vec{\mu}_{\textrm{n}}\cdot\vec{B}^*$.
The magnetic moment of the neutron is given by   $\vec{\mu}_{\textrm{n} }=\gamma_{\textrm{n}} \hbar \vec{s}$, where $\gamma_{\textrm{n}}  =- 2\pi \cdot 29.164 693 1(69) $~MHz/T is the neutron gyromagnetic ratio \cite{Greene}. In comparison, the corresponding magnetic field  due to the nuclear spin polarization $\vec{B}_{\textrm{mag}} = \mu_0 \frac{\hbar \gamma }{2} \rho  \vec{P}$ is typically three to four orders of magnitude weaker than the pseudo-magnetic field in  Eq.\ (\ref{eq.pseudoB}), where $\mu_0$ is the vacuum magnetic permeability and $\gamma$ is the gyromagnetic ratio of a spin-$\frac{1}{2}$ nucleus.
Analogous to a magnetic field, the pseudo-magnetic field induces a shift in the neutron Larmor precession frequency of
\begin{equation}
\begin{aligned}
\Delta f_n=-\frac{\gamma_{\textrm{n}}}{2\pi}B^{*}=\frac{2 \hbar}{m_{\textrm{n}}  }   \, b_{\textrm{i}} \rho P \sqrt{\frac{I}{I+1}}.
\end{aligned}
\label{eq.Freqbi}
\end{equation}
Employing this formula, $b_{\textrm{i}}$ can be extracted from a measurement of the neutron spin precession frequency. However, this requires absolute knowledge of the product of the nuclear number density and the degree of nuclear spin polarization.\\
In the presented study of mercury, the interaction occurs with an atomic gas where the atoms are not bound together. The relevant measured property is therefore the free scattering length $a_{\textrm{i}}$, which is related to the bound scattering length via
\begin{equation}
\begin{aligned}
b_{\textrm{i}} =\frac{M + 1}{M} \, a_{\textrm{i}} ,
\end{aligned}
\label{eq.biFromAi}
\end{equation}
with $M$ being the atomic mass number. For $^{199}$Hg with $M=199$, this yields $b_i=1.005 \, a_{\textrm{i}}$.


\begin{figure*}[htbp]
	\centering
	\includegraphics[width=0.95\linewidth]{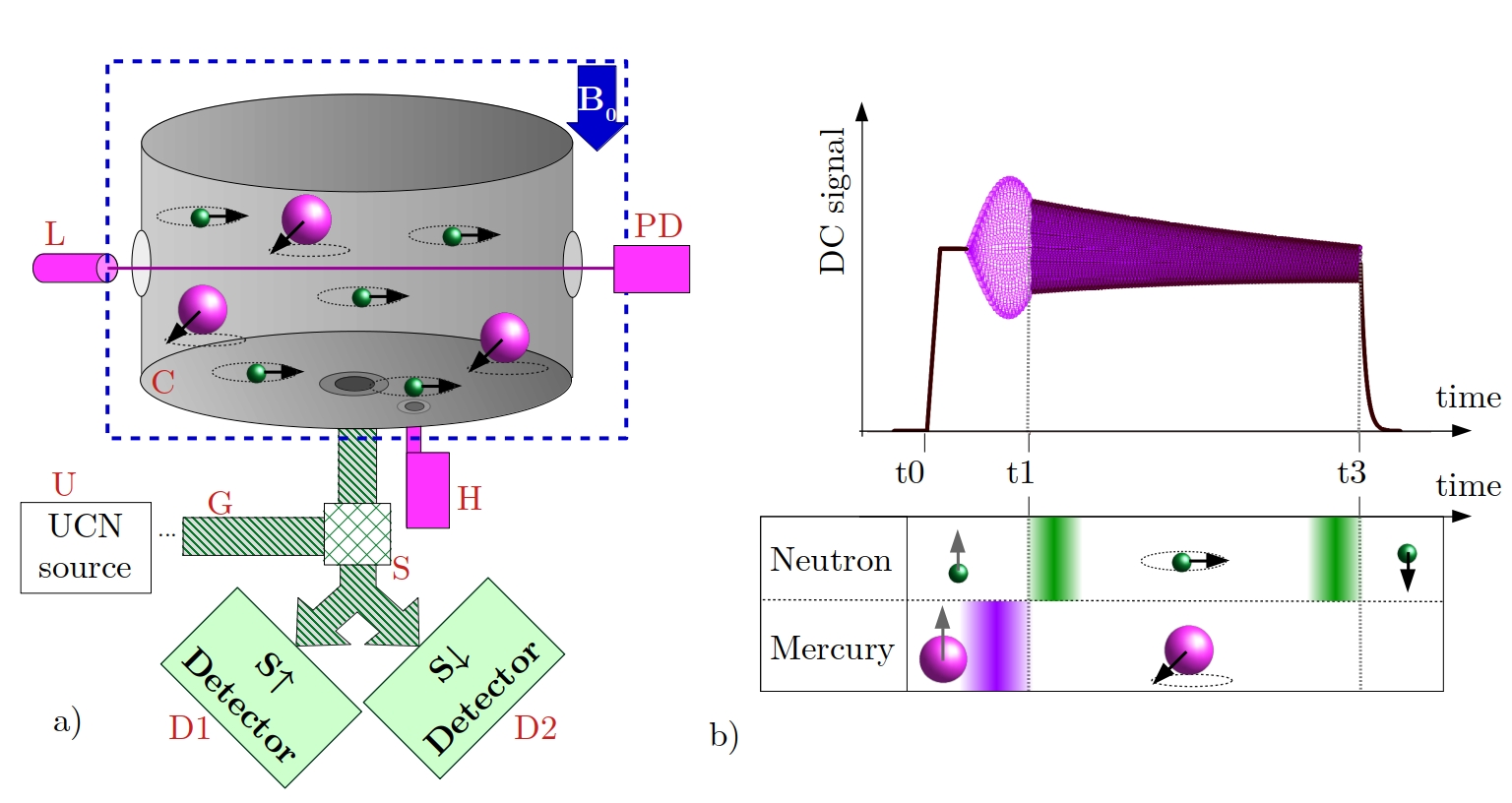}
\caption{(Color online) a) A sketch of the Ramsey  apparatus to measure the neutron EDM at the Paul Scherrer Institute. The small green spheres represent the neutrons and the large pink spheres the mercury atoms. The incoming ultracold neutrons (UCN) are transported from the source (U) to the switch (S), which guides them to the precession chamber (C) via neutron guides (G). After a certain storage time, the UCN are eventually transported from the precession chamber to the switch, and the detectors for spin-up and spin-down neutrons (D1, D2). The mercury system, indicated in pink colors, consists of the source (H) and a polarization cell situated below the precession chamber, the read-out laser (L) on one side, and a photo-detector (PD) on the opposite side. The blue dashed line represents the volume where the magnetic field $\lvert \vec{B}_0\rvert \approx 1\,$\textmu T is applied, and the blue arrow represents its orientation. This volume is shielded from the ambient magnetic field by four cylindrical layers of mu-metal, not shown for clarity.  The neutrons and mercury atoms are represented with their spin (black arrow) during the free precession. The coils used to perform the spin flips are not depicted for clarity.  
b) A simplified representation of the DC signal recorded by the photo-detector (PD) for one measurement cycle as a function of time.  The characteristic times $t_0$, $t_1$, and $t_3$ are related to the measurement sequence of the mercury and correspond to the points in time when the chamber starts to be filled with mercury, when the free precession of the mercury spins starts, thus, creating an associated oscillating signal, and when the emptying process of the chamber begins, respectively. 
Below the spin states of the mercury nuclei and the neutrons are shown when the neutron Ramsey technique is exactly on resonance and for a $\frac{3}{4}\pi$ mercury spin flip configuration. 
The spin flip pulses are indicated by the color shaded areas in the diagram. The duration of the mercury spin flip pulses are listed in Table~\ref{Table:Batches} and the neutron spin flips pulses have a length of $T_{SF}=2$~s. 
}
	\label{Fig:apparatus}
\end{figure*} 

\section{The nEDM Ramsey Apparatus}

For the incoherent scattering length measurement of $^{199}$Hg, the aforementioned nEDM apparatus was employed. Originally, it was designed for the dedicated search for a neutron electric dipole moment using a mercury co-magnetometer \cite{BAKER2014184,apparatusPSI,EDM2020,PhysRevD.92.092003}. 
An overview of the relevant components of the apparatus is given in Fig.\ \ref{Fig:apparatus}a.
The UCN source (U) at the Paul Scherrer Institute provides neutrons in pulses of 8~s length approximately every five minutes \cite{ANGHEL2009272,UCNSource,Bison2022,10.21468/SciPostPhysProc.5.004}. The neutrons are transported through UCN-reflecting vacuum guides (G) to a mechanical switch (S) which, when set to the filling position, connects the guides to the precession chamber (C). In the emptying position, the switch connects the chamber to the neutron spin analyzers and detectors (D1, D2). 
The precession chamber is a cylindrical neutron storage container with an inner diameter of 470~mm and a height of 120~mm, formed by a bottom and a top aluminum electrode and a polystyrene insulator ring. 
The electrodes are coated with diamond-like carbon and the inside of the insulator ring is coated with deuterated polystyrene to ensure optimum UCN storage properties \cite{PhysRevC.76.044001,BODEK2008222}. 
The chamber is placed in a vacuum tank inside a multi-layer magnetic shield with an internal set of coils generating an homogeneous vertical magnetic field $\lvert \vec{B}_0 \rvert \approx 1$~\textmu T along the cylinder axis of the precession chamber. This field is monitored during the measurement procedure by optically detected nuclear magnetic resonance of spin polarized mercury atoms present in the same volume as the UCN. 
This allows correcting for magnetic field drifts potentially occurring during the neutron spin precession frequency measurements \cite{Afach1112015}.\\ 
The mercury atoms are continuously produced in the source (H) situated below the precession chamber. The source thermo-dissociates HgO with a highly enriched $^{199}$Hg content of about 90\% and sub-percent residuals of $^{201}$Hg 
\cite{FertlThesis}.\footnote{There are two stable fermionic isotopes of mercury, $^{199}$Hg ($I=1/2$) and $^{201}$Hg ($I=3/2$), with natural abundances of approximately 17\% and 13\%, respectively. The residual impurities of the source do not affect the polarization measurement as the read-out laser is tuned to the specific optical transition wavelength of $^{199}$Hg. Moreover, the gyromagnetic ratios of the two spin species are different by a factor of about 0.37, such that their Larmor resonance frequencies do not overlap and the $^{201}$Hg nuclei are not affected by spin flip pulses \cite{PhysRevC.7.2065}.
In addition, no sizable $^{201}$Hg polarization has ever been observed in the apparatus, since the $I=3/2$ allows for electric quadrupole coupling leading to very short spin relaxation times.
}
The operation temperature of the source $T_{\textrm{Hg}}$ influences the atomic mercury gas density, i.e.\ the higher the temperature, the more mercury is released.
In the presented configuration, a right-handed circularly polarized light polarizes the mercury atoms in a polarization cell connected to the source, aligning their nuclear spins anti-parallel to the direction of the main magnetic field $\vec{B}_0$. 
%
%
%
After this process, the atoms diffuse further into the precession chamber through a mechanical valve opening.
The spin orientation and density of the mercury atoms are continuously monitored via a light absorption measurement. 
Ultraviolet light, originating from a laser (L) traversing the precession chamber, is tuned to the transition of the electronic states of ${}^{199}$Hg, $6^{1}S_0 \rightarrow 6^3 P_1$. The laser beam passes through the mercury gas in the chamber to  a photo-detector (PD) on the opposite side. It consists of a photo-multiplier tube coupled to a low-noise current amplifier, generating a signal proportional to the transmitted light intensity.  
The data acquisition system connected to the photo-detector records two separate voltage signals: a DC-coupled signal with a sampling rate of up to 50~Hz, which is presented in Fig.\ \ref{Fig:apparatus}b, and an AC-coupled signal sampled at 100~Hz which first passes through a band-pass filter centered at approximately 8~Hz, matching the Larmor frequency of $^{199}$Hg. 
A detailed description of the mercury magnetometry system is presented in Ref.\ \cite{PSI_HgcoMag}.\\
To induce flips of the mercury and neutron spins, two sets of coils were installed. 
A first set surrounding the vacuum tank combined with a saddle coil is used to produce a circulatory rotating field transverse to $\vec{B}_0$.  
A second set of coils produces resonant linear oscillating magnetic fields in the plane perpendicular to $\vec{B}_0$. 
The amplitude, duration, and frequency of two oscillating field pulses are adjusted to sequentially flip the two spin species. 
The frequencies of the flip pulses are  separated far enough such that they do not influence the other species, roughly 8~Hz for $^{199}$Hg and 29~Hz for neutrons at a field of 1~\textmu T.

\section{Experimental method}
\label{sec:expmethod}


The incoherent scattering length of ${}^{199}$Hg is measured with the nEDM apparatus. The mercury co-magnetometer acts as the source of polarized nuclei, which cause a corresponding pseudo-magnetic field. 
The measurement employs Ramsey's technique of separated oscillatory fields adapted to UCN \cite{ramsey1949,Ramsey,RAMSEY1986223}.
The neutron spin precession frequency is determined for different densities and polarization values of the mercury gas, which is co-inhabiting the precession chamber. 
A shift of the precession frequency caused by the incoherent scattering length is maximum if the nuclear mercury spins are in pure eigenstates with respect to the external magnetic field $\vec{B}_0$, i.e.\ parallel or anti-parallel, respectively.
However, the visibility of the co-magnetometer signal  is maximum for the mercury spins precessing in the plane orthogonal to the magnetic field. 
A compromise was chosen by preparing the mercury spins in a superposition state corresponding to a $\frac{3}{4}\pi$ or $\frac{7}{4}\pi$  spin flip from the initial spin state. This reduces the strength of the incoherent scattering length effect and of the visibility of the co-magnetometer signal each by a factor $\sqrt{2}$. \\
The neutron spin precession frequency is determined from a Ramsey pattern obtained from several so-called cycles. One such cycle consists of the following set of steps leading to the determination of the final spin state of the neutrons:



\begin{itemize}
\item Filling: At the beginning of each cycle,
the UCN are first filled into the precession chamber and, in a second step,  the mercury atoms are introduced.
Initially, the spins of the neutrons and mercury nuclei are polarized along the magnetic field axis with the
nuclear spin of the mercury being  anti-parallel to the main magnetic field direction.

\item Spin-flips: A first circular rotating magnetic field pulse is applied to flip the mercury spins. Its frequency is set to match the mercury Larmor frequency, $f_{\textrm{Hg}}$, of about $8$~Hz at $1 \textrm{ \textmu T}$. Its duration depends on the rotating field amplitude and the superposition state aimed for, i.e.\ $\frac{3}{4}\pi$ or $\frac{7}{4}\pi$ spin flip in Table~\ref{Table:Batches}. 
Subsequently, a linear oscillating magnetic field pulse is applied to flip the UCN spins by $\frac{\pi}{2}$  into the plane orthogonal to the magnetic field $\vec{B}_0$. Its frequency, $f_{\textrm{RF}}$,  is close to the neutron Larmor frequency of approximately $29$~Hz and its duration is $T_{SF}=2\,\textrm{s}$.

\item Free precession: The UCN and mercury spins precess freely in the magnetic field $\vec{B}_0$. The free precession time of the neutrons lasts for $T_0=180$~s. 

\item Spin-flip: After the free precession, another identical and phase-locked linear oscillating magnetic field $\frac{\pi}{2}$-pulse is applied to the UCN spins. 

\item Emptying: The UCN shutter is opened to empty the precession chamber from the mercury atoms and the UCN. The latter are guided to the spin analyzer and detectors \cite{Ban112016,Afach22222015}. The analyzer delivers the neutrons  to the detector D1 or D2 according to their final spin state.  From the time-integrated counts in each detector, a counting-asymmetry is computed 
\begin{equation}
A=\frac{N_{1}-N_{2}}{N_{1}+N_{2}},
 \label{Asym}
 \end{equation}
 where $N_{1}$ and $N_2$ correspond to the neutron counts in detector D1 and D2, respectively. 
 During the emptying process, the residual Hg vapor is pumped out of the precession chamber.
\end{itemize}
Such a cycle is repeated in total twelve times at four different $f_{\textrm{RF}}$ settings to obtain one Ramsey pattern for a given mercury spin configuration, i.e.\  (a) $\frac{3}{4}\pi$  or (b) $\frac{7}{4}\pi$ spin flip. The four frequencies are chosen to be located in the steepest region of the central Ramsey fringe (compare Fig.\ \ref{Fig:CosFit}).
Ramsey patterns are taken repeatedly with alternating mercury spin configurations in a Thue–Morse sequence (abbabaab), with otherwise ideally stable mercury density and polarization conditions. This measurement sequence helps to compensate for potential drifts of the magnetic field. 
Two subsequent Ramsey patterns (ab or ba) form a polarization group.\footnote{When a complete Ramsey pattern has to be rejected from the data analysis, its accompanying Ramsey pattern is incorporated into the following polarization group. See Sec.\ \ref{DataSelection} and Table~\ref{Table:Batches}.
}
The data-taking campaign was performed over several days in three measurement batches summarized in Table~\ref{Table:Batches}. 
For instance, in one of the batches, the temperature of the mercury source was intentionally lowered to achieve a different mercury gas density. The total net data-taking duration was approximately 60~hours.




\begin{table}[tbp]
\begin{center}
\begin{tabular}{l| >{\centering}m{1cm} >{\centering}m{1cm} >{\centering}m{1cm} | c}
	\hline
	\hline
Batch number	  & 1  & 2  & 3 & total \\ 
\hline
No.\ of cycles	& 240  & 228  & 204 & 672 \\[1.5pt]
No.\ of Ramsey patterns & 20  & 19  & 17 &56  \\[1.5pt]
No.\ of polarization groups & 10 & 8 & 8 & 26\\[1.5pt]  
Source temperature $T_{\textrm{Hg}}$ (°C)  & 225  & 225  & 210  &  \\[1.5pt]
Hg signal sampling rate (Hz)  & 10  & 10  & 50  &  \\[1.5pt]
Hg $\frac{3}{4}\pi$-spin flip pulse length (s)  & 1.5   & 3.0  & 3.0   &  \\[1.5pt]
Hg $\frac{7}{4}\pi$-spin flip pulse length (s)  & 3.5   & 7.0  & 7.0   &  \\[1.5pt]
\multirow{2}{*}{$\langle \lvert \rho P \rvert \rangle$ ($10^{15}$m$^{-3}$) } & 7.1   & 6.4 & 5.13    &  \\[-2pt]
 &  $\pm$0.1 &  $\pm$0.1   & $\pm$0.07   &  \\
 \hline
\hline
\end{tabular} 
\end{center}
\caption{Parameters of the different batches with the number of cycles, Ramsey patterns, and groups with opposite polarization taken at two mercury source temperatures with two different sampling rates of the DC-coupled mercury photo-detector signal. In addition, the duration of the corresponding Hg spin flip pulses and the average value for the product of number density and polarization is given in absolute values.
}
\label{Table:Batches}
\end{table}

\section{Data Analysis}
\label{sec:analysis}

\begin{figure}
	\centering
\includegraphics[width=0.95\linewidth]{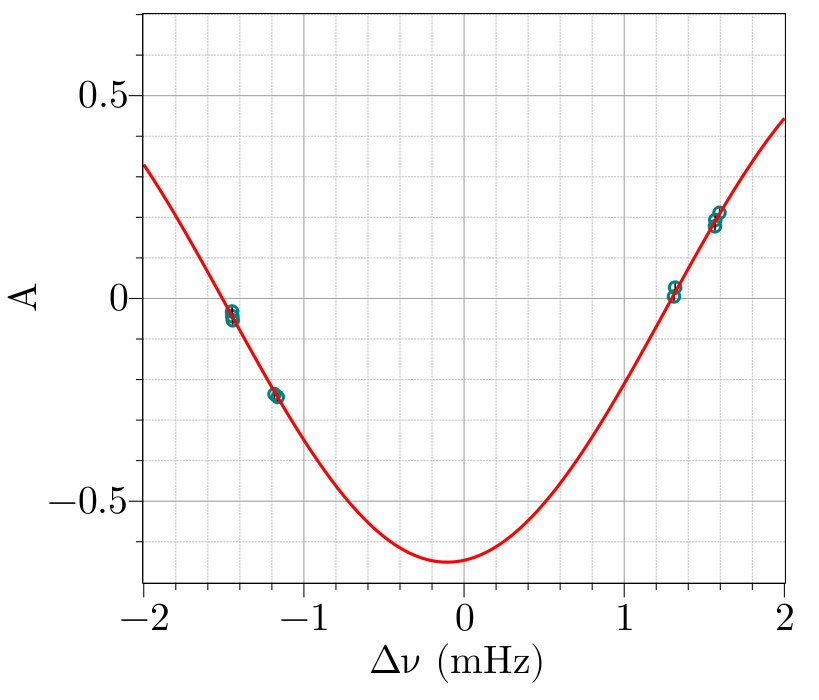}
	\caption{Example of asymmetry $A$ vs. $\Delta \nu$  plot from batch~3. Here, only ten data points are used for the fit with Eq.\ (\ref{eq.sinFit}) (red curve) due to the selection criteria described in Sec.\ \ref{DataSelection}.
    }
	\label{Fig:CosFit}
\end{figure}

Using Eq.~(\ref{eq.Freqbi}) and conveniently approximating here $a_i \approx b_i$  for mercury in Eq.~(\ref{eq.biFromAi}), one can derive the bound scattering length to be
\begin{equation}
b_\textrm{i} = \frac{m_{\textrm{n}}}{2 \hbar}  \sqrt{\frac{I+1}{I}} \frac{ \Delta f_\textrm{n}}{\rho\, P},
 \label{eq.ai}
 \end{equation} 
where $\Delta f_n$ is the shift of the neutron Larmor precession frequency due to the pseudo-magnetic interaction. 
%
The goal of this experiment is to evaluate this frequency shift by performing a relative measurement between the two mercury spin configurations described in Sec.\ \ref{sec:expmethod}.
This involves analyzing data from each polarization group in order to acquire the slope $\partial (\Delta f_{\textrm{n}})/\partial(\rho P)$ via a linear regression. As $\Delta f_\textrm{n}$ and $\rho P$  are deduced from two distinct measurements (neutron Ramsey data and mercury precession data), the determination of these two values can be done independently. \\
%
The neutron precession frequency shift $\Delta f_\textrm{n} =f_\textrm{n}-|\gamma_{\textrm{n}}/\gamma_{\textrm{Hg}}|  f_{\textrm{Hg}}$ is obtained for each individual cycle, where $f_{\textrm{n}}$ is the actual neutron Larmor precession frequency, $f_{\textrm{Hg}}$ is the Larmor precession frequency of mercury, and $\gamma_{\textrm{Hg}} = 2\pi \cdot 7.590 118 (13)$~MHz/T is the gyromagnetic ratio of $^{199}$Hg \cite{Hgmag,AFACH2014128}.\footnote{If a drift in the global magnetic field affects the neutrons and the mercury spin equally,  the resulting frequency shift does not alter $\Delta f_{n}$.   As the pseudo-magnetic field from the incoherent scattering length only affects the neutron spin, its effect remains detectable in $\Delta f_{n}$.} 
%
 %
During the measurement of a Ramsey pattern, $f_{\textrm{RF}}$ is set in the range of the central fringe as depicted in Fig.~\ref{Fig:CosFit}. The corresponding asymmetry, defined in Eq.\ (\ref{Asym}), has in first approximation a cosine behavior and the data can be fitted as a function of $\Delta \nu$ by 
\begin{equation}
 g(\Delta \nu) = A_0 - |\alpha| \cdot \cos \left[ 2\pi T \, (\Delta \nu+\Delta f_{\textrm{n}}) \right],
 \label{eq.sinFit}
 \end{equation}
 with
 \begin{equation}
\Delta \nu=\left|\frac{\gamma_{\textrm{n}}}{\gamma_{\textrm{Hg}}}\right|f_{\textrm{Hg}}-f_{\textrm{RF}},
 \label{nu}
 \end{equation}
where  $\alpha$, $A_0$ and $\Delta f_{\textrm{n}}$ are the fit parameters and $T= T_0+ \frac{4  T_{SF} }{\pi} \approx 182.5\, \textrm{s}$ is the effective interaction time \cite{Piegsa2008b}. The latter takes the entire neutron spin precession time $T_0$ with the neutron spin flip pulse duration $T_{SF}$ into account.
Assuming the values of $A_0$ and $\alpha$ to be constant over the course of one Ramsey pattern measurement time, one can compute the neutron frequency shift for each individual cycle with
%
  \begin{equation}
 \Delta f_{\textrm{n},j} = \frac {\sgn (\Delta \nu_j)}{2\pi T} \; \arccos \left(\frac{{A_0}-A_j } { | \alpha | }\right)-  {\Delta \nu_j} ,
 \label{delta f}
 \end{equation}
where $j$ is an index running over all cycles of one Ramsey pattern, $A_j$  the asymmetry value of cycle $j$, and $\sgn (\Delta \nu_j)$ gives the sign of $\Delta \nu_j$.  This sign is used in Eq.~(\ref{delta f}) to expand the range of the $\arccos$-function from [0,$\pi$] to [-$\pi$,$\pi$]. \\

\begin{figure}
	\centering
	\includegraphics[width=0.95\linewidth]{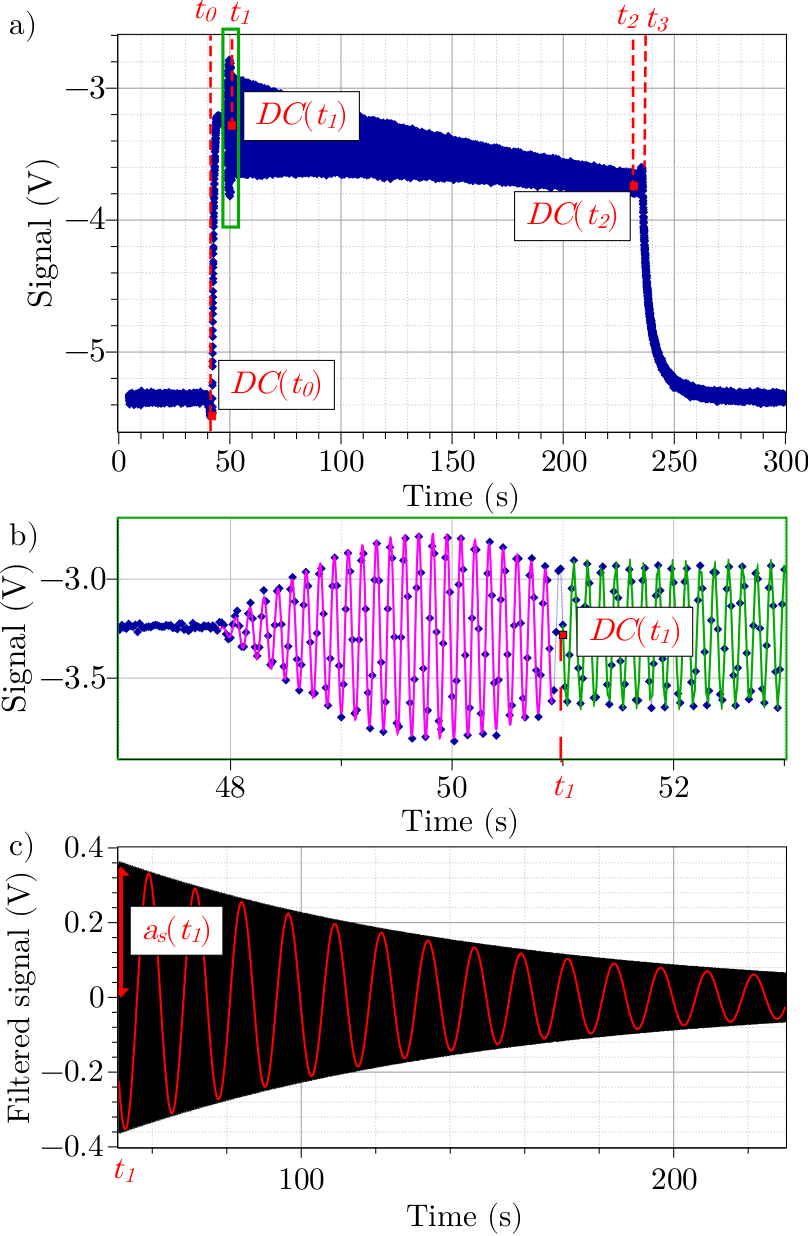}
	\caption{a) Example of a DC-coupled signal of the ${}^{199}$Hg co-magnetometer for a $\frac{3}{4}\pi$ spin configuration. The $DC(t)$ values are represented by red squared dots for the times $t_0$, $t_1$, and $t_2$.  b) Detailed view of the green framed region in Fig.\ 3a. The first part of the signal corresponds to the mercury spin flip and is fitted by a double sinusoidal function indicated in pink (fast sinusoidal oscillation multiplied with a sinusoidal envelope). The value of $DC(t_1)$ is determined from this fit. The green line is a guide for the eyes to see the oscillating signal caused by the free precession of the mercury spins.  c) An example of the AC-coupled filtered signal between $t_1$ and $t_3$ (black curve) of the photo-detector of the same cycle.  The red curve is a representation of the decay sinusoidal function used to fit the data but with its period scaled by a factor of 100 so that the oscillations become visible. For this cycle, the amplitude is $a_s(t_1)=0.36\,$V, the running averages of the light signal are $DC(t_0)=-5.48\,$V, $DC(t_1)=-3.28\,$V, and the decay time constants are $T_2 = 101\,$s and $T_3=810\,$s.  
    }
	\label{Fig:PMT}
\end{figure}

The product of the mercury number density and polarization inside the precession chamber is calculated from the mercury light absorption signal recorded with the photo-detector. An example of such a measurement is shown in Fig.~\ref{Fig:PMT}.  The transmitted light intensity passing through the chamber decreases when the number of mercury atoms increases due to the absorption from mercury atoms. 
Hence, the running average of the light signal at a time $t$, denoted $DC(t)$, is a measure for the mercury density.
The free precession of the mercury spins cause a modulation of the read-out laser intensity reaching the photo-detector due to the spin-dependent absorption cross-section. As a consequence, the amplitude of the oscillation $a_s(t)$ is proportional to the density of polarized mercury atoms in the vapor.
%
The general expression for the  product of the mercury number density and polarization is given by \cite{FertlThesis,PSI_HgcoMag} 
 \begin{equation}
 \rho P(t)=  \frac{1}{ \sigma L } \, \textrm{arcsinh}\left(\frac{a_s(t)}{DC(t)}\right),
 \label{eq:DPG}
 \end{equation}
 where $\sigma = (1.97 \pm 0.04) \times 10^{-13}\,\mathrm{cm}^{2}$ is the theoretical spin-independent light absorption cross-section of mercury derived in Appendix A, and $L=470 \,\mathrm{mm}$ is the diameter of the precession chamber. In Fig.\ \ref{Fig:PMT}, an exponential decrease in $DC(t)$ and $a_s(t)$ between $t_1\approx 51\, $s and $t_3 \approx 230\,$s is observed. This is due to the leakage of mercury vapor out of the chamber in the surrounding vacuum tank and  the depolarization of the mercury spins due to wall collisions, respectively. This behavior can be parametrized as
 \begin{equation}
DC(t)=[DC(t_1)-DC(t_0)]\cdot\textrm{exp}\left(-\frac{t-t_1}{T_{3}} \right)+DC(t_0)
 \label{eq:DC2Decay}
 \end{equation}
 and
  \begin{equation}
a_s(t)=a_s(t_1) \cdot \textrm{exp}\left( - \frac{t-t_1}{T_2} \right),
 \label{eq:asDecay}
 \end{equation}
where $t_0$ is the start time when the chamber is filled with mercury, $t_1$ is the time at which the mercury flipping pulse ends and the free precession starts, $T_2$ is the depolarization time constant, and $T_{3}$ is the leakage time constant computed from
\begin{equation}
T_{3}= \frac{t_2-t_1}{\textrm{ln}\left(\frac{DC(t_1)-DC(t_0)}{DC(t_2)-DC(t_0)}\right)},
 \label{eq:Tleak}
  \end{equation}
where $t_2$ is a time shortly before the time $t_3$ at which the emptying process of the precession chamber begins.
For a better understanding, $t_0$, $t_1$, $t_2$, and $t_3$ are indicated in Fig.\ \ref{Fig:apparatus}b and Fig.\ \ref{Fig:PMT}a and b.
The values of $DC(t)$ are determined from the DC-coupled signal shown in Fig.\ \ref{Fig:PMT}a and b, whereas $a_s(t_1)$ and $T_2$ are the result of a fit to the AC-coupled filtered signal, represented in Fig.\ \ref{Fig:PMT}c, avoiding the distortion of the oscillations by the shape of the DC-signal. 
Note, a drop in the DC-signal is visible just before $t_0$ due to the closing of the mechanical neutron shutter. Since this movement leads to a small rotation of the precession chamber, the alignment between the chamber and the read-out laser slightly changes. This specific alignment is then maintained during the entire free precession measurement. 
The opposite effect is  visible when the shutter opens again to empty out the chamber at $t_3$. 
Because of this change of alignment, $DC(t_0)$ and $DC(t_2)$ are measured after and before the shutter movement, respectively.\\
By inserting Eq.\ (\ref{eq:DC2Decay}) and Eq.\ (\ref{eq:asDecay}) into Eq.\ (\ref{eq:DPG}) and performing a time integration, one can determine an effective product of the mercury number density and polarization averaged over the interaction period $T$
 \begin{equation}
 \rho P= \frac{1}{ \sigma L T} \int_{t_1}^{t_1+T} \textrm{arcsinh}\left(\frac{a_s(t)}{DC(t)}\right) dt.
 \label{eq:DPeff}
  \end{equation}
The sign of the product is determined according to the mercury spin configuration: 
in the $\frac{3}{4}\pi$ configuration, the polarization vector points in the direction of the main magnetic field $\vec{B}_0$, therefore, $\rho P$ has a positive value, while in the $\frac{7}{4}\pi$ configuration the polarization is oriented anti-parallel to the main magnetic field and $\rho P$ is negative.  




\section{Data selection criteria \& Results}
\label{DataSelection}

\begin{figure*}[t]
    \includegraphics[width=0.95\linewidth]{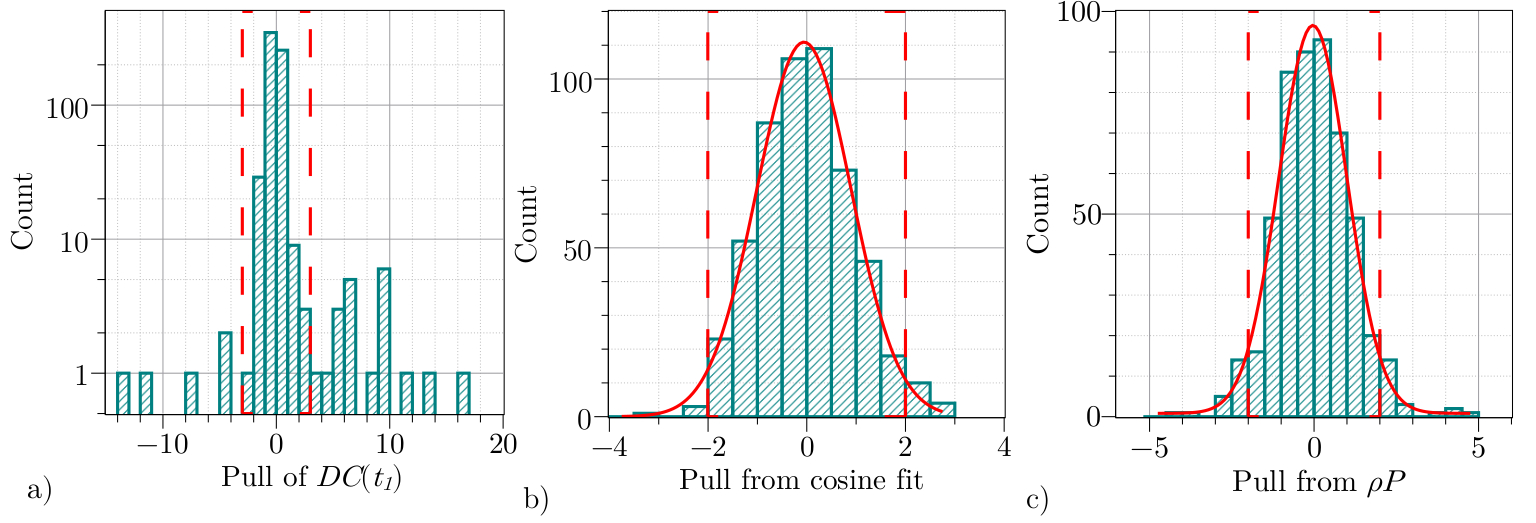}
    \caption{ a) Pull distribution of $DC(t_1)$   as defined in Eq.\ (\ref{eq:PullDC}). The red dashed lines represent the exclusion limits. Note, the vertical axis is in log-scale for a better visibility. b) Pull distribution from Eq.\ (\ref{eq:PullCos}). The red dashed lines represent the exclusion limits and the continuous red line is a Gaussian fit to the distribution. c) Pull distribution of the product of the mercury number density and polarization  as defined in Eq.\ (\ref{eq:PullrhoP}). The red dashed lines represent the exclusion limits and the continuous red line is again a Gaussian fit.}
	\label{Fig:tsExclusion}
\end{figure*}

From an initial set with a total of 672 cycles, data were cut at three different stages of the analysis. \\
First, 24 cycles were removed due to exceptional events, e.g.\ an accidental interruption in the middle of a Ramsey pattern etc. \\
Then, during the data extraction, a set of criteria based on three parameters is applied to the data, described in the following. 
The first parameter is the waiting time: the time between the end of the previous cycle and the beginning of the subsequent one.  Fluctuations can occur as the waiting time depends on the time between pulses of the UCN source which is not fixed. 
For a waiting time shorter than $10\,$s, the mercury source has not yet built up enough density, and for a waiting time longer than $21\,$s, the density is too high. 
Because of this rough criteria, 90 additional cycles were excluded. 
The second parameter is the value of $DC(t_1)$ of each cycle. As stated in the previous section, this parameter is also directly related to the density of mercury inside the chamber.  
The acceptable value range of $DC(t_1)$  is determined for each batch individually from their mean values $\langle DC(t_1) \rangle$ and standard deviations $\sigma(DC(t_1))$, both calculated after the initial cuts in the data. Additionally, seven cycles outside a three-sigma range were rejected.  The cut on the full data set with the pull distribution is presented in Fig.\ \ref{Fig:tsExclusion}a, where the pull value is defined by 
 \begin{equation}
P_{DC}=\frac{\langle DC(t_1) \rangle -DC(t_1)}{\sigma(DC(t_1))}.
 \label{eq:PullDC}
  \end{equation}
The third parameter is neutron statistics. Low statistic measurements with fewer than 4000 detected neutrons were removed, in comparison to an average neutron count of about 8000 per cycle. This leads to a rejection of 10 additional cycles.\\
Lastly, before the final stage of the analysis, a further set of criteria is applied to each Ramsey pattern. 
From the determination of the neutron frequency shift, the stability of the fit parameters  over a Ramsey pattern is estimated from the pull value of each cycle. The latter is defined by the difference between the value of the fit function Eq.\ (\ref{eq.sinFit}) and the data point, normalized by the uncertainty 
 \begin{equation}
P_{cos}=\frac{ g(\Delta \nu_j)-A_j}{\sigma (A_j)}.
 \label{eq:PullCos}
  \end{equation}
  Eighteen cycles further away than two sigma from their fit function were rejected during the frequency shift determination, as depicted in Fig. \ref{Fig:tsExclusion}b.
Moreover, from the determination of the mercury
number density and polarization of mercury, the stability of the product over each Ramsey pattern is estimated from the deviation between the data and the mean value $\langle\rho P \rangle$. Thirty-three cycles outside a two-sigma range are rejected. The cut on the data  is represented on the pull distribution in Fig.\  \ref{Fig:tsExclusion}c where the pull value is defined as
\begin{equation}
P_{\rho P}=\frac{\langle \rho P \rangle -(\rho P)}{\sigma(\rho P)}.
 \label{eq:PullrhoP}
  \end{equation}\\
%
%
%
%
%
%
%

\begin{figure}[!ht]
	\centering
	\includegraphics[width=0.95\linewidth]{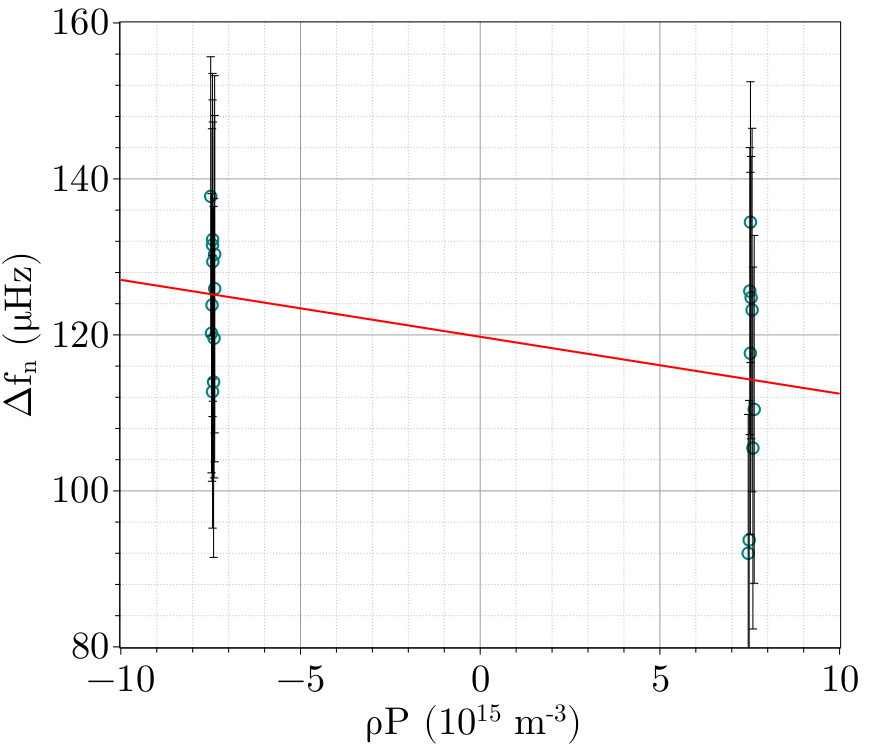}
	\caption{ An example of linear regression for one polarization group using the same group of data as in the previous figures \ref{Fig:CosFit} and \ref{Fig:PMT}.
    A positive (negative) value of $\rho P$ corresponds to the $\frac{3}{4}\pi$ ($\frac{7}{4}\pi$) mercury spin configuration. 
    } 
	\label{Fig:LinReg}
\end{figure}

\begin{figure}[!ht]
	\centering
	\includegraphics[width=0.95\linewidth]{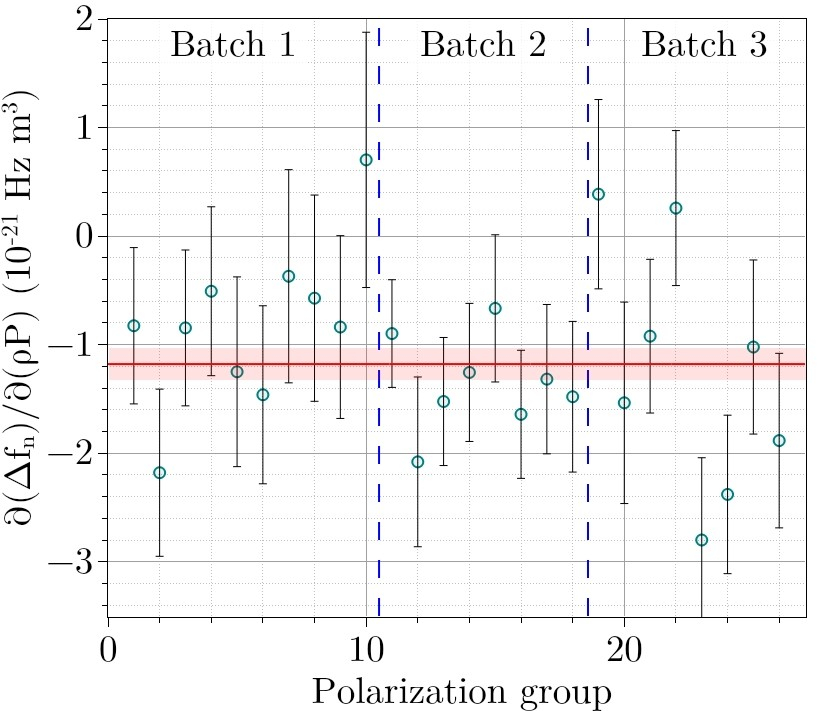}
	\caption{
 Results for the slopes of the linear regressions between $ \Delta f_\textrm{n}$ and $\rho P$ for each polarization group, i.e.\ successive Ramsey patterns with opposite mercury polarization. The data of batches 1, 2, and 3 are delimited by vertical dashed lines. The red horizontal line and band represent the weighted mean and standard deviation of  $\langle\partial( \Delta f_\textrm{n})/\partial( \rho P)\rangle ~= (-1.18\pm 0.15)\times 10 ^{-21} \,\textrm{Hz}\, \textrm{m}^3$. The residual sum of squares is 25.8 for 26 data points leading to a corresponding reduced $\chi^2$ of approximately 1.03.
} 
	\label{Fig:ConstFit}
\end{figure}

Ultimately, a total of 490 cycles out of 672 are used in the final stage of the analysis to determine the slope of the neutron frequency shift as a function of the product of the mercury number density and polarization  $\partial (\Delta f_{\textrm{n}})/\partial(\rho P)$. 
An example of one such analysis for one polarization group is shown in Fig.\ \ref{Fig:LinReg}, where the slope is deduced from a linear regression.
The full analysis for all batches and corresponding polarization groups is summarized in  Fig.\ \ref{Fig:ConstFit}. 
A weighted average yields a value for the slope of 
  \begin{equation}
  \langle \partial(\Delta f_\textrm{n})/\partial( \rho P)\rangle ~= (-1.18\pm 0.15) \times 10 ^{-21}\,\textrm{Hz}\, \textrm{m}^3. 
 \label{eq.slope}
 \end{equation}
By inserting  this value into Eq.\ (\ref{eq.ai}), one obtains
  \begin{equation}
b_{\textrm{i}}=( - 16.2  \pm 2.0) \, \textrm{fm.} 
 \label{biResult}
 \end{equation}
This value is in agreement with the literature value $|b_{\textrm{i},lit}|=(15.5\pm 0.8)\,\textrm{fm}$ and the negative sign of the neutron incoherent scattering length of ${}^{199}$Hg was determined.



\section{Conclusion}

In this article, we have presented the direct measurement and analysis of the mercury incoherent scattering length using the neutron Ramsey apparatus of the nEDM collaboration located at the Paul Scherrer Institute. The result of $b_{\textrm{i}}=( - 16.2  \pm 2.0) \, \textrm{fm}$ is in agreement with the absolute value found in the literature. 
It also provides the so far unknown sign of the scattering length, which can be used to correct the value of the last neutron EDM measurement and diminish its systematic error budget \cite{EDM2020}. 
The effect of the incoherent scattering length is of importance for any future neutron electric dipole moment experiments using mercury as a co-magnetometer \cite{TriumfXecoMag,Ayres2021}. These experiments should therefore anticipate the determination of the polarization and the density of the mercury during their measurements. 
A similar situation will also appear with the use of a  ${}^{129}$Xe co-magnetometer for which the incoherent scattering length has been measured recently \cite{lu2023measurement}.

\begin{acknowledgments}
We acknowledge the excellent support provided by the PSI technical groups and by various services of the collaborating universities and research laboratories.\\
The Swiss research groups acknowledge financial support from the Swiss National Science Foundation through project nos.\ 163663, 181996, and 215185 (University of Bern), nos.\ 117696, 126562, 137664, 144473, 157079, 169596, 172626, 178951, and 188700 (Paul Scherrer Institute), and nos.\ 162574, 172639, 200441, and 10003932 (ETH Zürich).
University of Bern also acknowledges the support via
the ERC Project no.\ 715031-BEAM-EDM.
Contributions of the
Sussex group have been made possible via the School of Mathematical and Physical Sciences at the University of Sussex, as well as the STFC under grant ST/S000798/1 and ST/N000307/1.
This project was supported by the Research Foundation Flanders (FWO) with grants G.0375.09N and G0D0421N, and by Project GOA/2010/10 of the KU Leuven.
LPC Caen and LPSC Grenoble acknowledge
the support of the French Agence Nationale de la Recherche
(ANR) under Reference nos.\ ANR-14-CE33-0007, ANR-09-BLAN-0046,  and the ERC Project no.\ 716651-NEDM.
%
%
The collaborators from the Jagiellonian University in Cracow wish to acknowledge support from the National Science Center, Poland, under grant no.\ 2020/37/B/ST2/02349 and also by the Minister of Education and Science under the agreement no.\ 2022/WK/07.
P.M.\ would like to acknowledge support from SERI FCS award 2015.0594.


\end{acknowledgments}

\appendix
\section{Light absorption cross-section of mercury}
In this appendix, the light absorption cross-section of mercury $\sigma$ employed in Eq.\ (\ref{eq:DPG}) is derived. Due to competing conventions several variable names have already been used previously in a different context in the main text of the article and should not be confused.\\

If an atom has two populated states $E_1$ and $E_2$, the population of these two states is ruled by: the spontaneous emission from $E_2$ to $E_1$, the induced emission from $E_2$ to $E_1$, and the induced absorption from $E_1$ to $E_2$. The probability of these interactions is defined by the Einstein coefficients $A$ for the spontaneous emission and $B_{12}$ ($B_{21}$) for the induced absorption  (induced emission). The probability of induced emission or absorption $d\mathcal{P}/dt$ is proportional to the density of light $\rho$ that stimulates it: 
\begin{subequations}
\begin{align}
\frac{d}{dt} \mathcal{P}_{21}^{\textrm{ind}}=B_{21}\rho,   \\
\frac{d}{dt} \mathcal{P}_{21}^{\textrm{spont}}=A,  \\   
\frac{d}{dt} \mathcal{P}_{12}^{\textrm{ind}}=B_{12} \rho. 
\end{align}
\label{eq:B12}
\end{subequations}

In equilibrium, the emission of light is balanced by its absorption,
\begin{equation}
N_2 \frac{d}{dt} \mathcal{P}_{21}^{\textrm{ind}} +  N_2 \frac{d}{dt} \mathcal{P}_{21}^{\textrm{spont}}=N_1 \frac{d}{dt} \mathcal{P}_{12}^{\textrm{ind}},
  \label{eq:XSequi}
\end{equation}
where $N_i$ is the population density of the state $i$. In equilibrium this follows a Boltzmann distribution:
 \begin{equation}
N_i=N \frac{g_i}{Z}e^{-E_i/kT},
\label{eq:boltz}
\end{equation}
  where $N$ is the total population density, $g_i$ the number of degenerate sub-levels, $Z$ a normalization factor, and $k$ is the Boltzmann constant.
 Using Eq.\ (\ref{eq:B12}) and Eq.\ (\ref{eq:boltz}), Eq.\ (\ref{eq:XSequi}) becomes 
 \begin{equation}
\rho = \frac{A/B_{21}}{\frac{g_1 B_{12}}{g_2 B_{21}}e^{h\nu_{21}/kT}-1},
\label{eq:XSSpecDensity}
\end{equation}
where $h \nu_{21} = E_2-E_1$. As this must follow Plank's law at all temperatures and frequencies, the Einstein coefficients have the relations:
\begin{subequations}
 \begin{align}
 B_{12}=\frac{g_2}{g_1}B_{21},  \\
 A=\frac{8 \pi h (\nu_{21})^3}{c^3}B_{21}. 
 \end{align}
\label{eq:EinstCoeffR}
\end{subequations}

The absorption can also be described by its cross-section $\sigma_0$ with 
\begin{equation}
B_{12}=\frac{c}{h \nu_{21}} \sigma_0.
\label{eq:XSandB12}
\end{equation}
Reversing this formula and expressing it using $A$, the cross-section becomes
\begin{equation}
\sigma_0=\frac{B_{12} h \nu_{21}}{c}=\frac{B_{21}\frac{g_2}{g_1} h \nu_{21}}{c}=\frac{g_2}{g_1}\frac{ A  }{ 8 \pi  }\lambda^2.
\label{eq:XSandA}
\end{equation}
For the Doppler broadened cross-section, the cross-section has a corrective factor of
\begin{equation}
\sigma (\nu) = \frac{\sigma_0}{ \sqrt{2\pi}  \Delta}\textrm{exp}\left (- \frac{(\nu-\nu_{21})^2}{2 \Delta^2} \right),
\label{eq:XSd}
\end{equation}
where $\Delta$ is the standard deviation of the Doppler width. 
As we consider the monochromatic case tuned to the correct frequency  $\nu=\nu_{21}$ the exponential term becomes one and the cross-section is 
\begin{equation}
\sigma= \frac{g_2}{g_1}\frac{ A  }{ 4  } \frac{ \lambda^2}{ (2\pi)^{3/2} \Delta},
\label{eq:XSc}
\end{equation}
where
\begin{equation}
\Delta=\sqrt{\frac{k T}{Mc^2}}   \nu_{21}.
\label{eq:XSb}
\end{equation}
In the case of the transition from the $6 {}^{1}S_0 $ state to the $6 {}^{3}P_1$ state for ${}^{199}\textrm{Hg}$, $\Delta=(436 \pm 4)\,\textrm{MHz}$ at room temperature $T=(293 \pm 5)\,\textrm{K}$ and $g_1=g_2$ as the hyperfine structure is well separated and  only the $F=1/2$ sub-levels are targeted. Then, with $A= [(119\pm 2)\,\textrm{ns}]^{-1} = (8.40 \pm 0.14)\,\textrm{MHz}$ and $\lambda= c/ \nu_{21} = 253.7\,\textrm{nm}$ \cite{DOIDGE1995209}, the light absorption cross-section is \mbox{$\sigma=(1.97 \pm 0.04)\times 10^{-13}\, \textrm{cm}^2 $}.

\nocite{*}

\bibliography{apssamp}

\end{document}